\documentclass{ws-ijmpa}

\begin{document}

%

\def\nocropmarks{\vskip5pt\phantom{cropmarks}}

\let\trimmarks\nocropmarks      

%

\markboth{Y.-J. Zhang, B.-Q. Ma \& L.-M. Yang} {Flavor Asymmetry
of Nucleon Sea from Detailed Balance}

%
\catchline{}{}{}
%

\setcounter{page}{1}

\title{
FLAVOR ASYMMETRY OF NUCLEON SEA FROM DETAILED BALANCE\footnote{
Talk presented at the Third Circum-Pan-Pacific Symposium on ``High
Energy Spin Physics", Oct.~8-13, 2001, Beijing, China. This work
is partially supported by the National Natural Science Foundation
of China. }}

\author{\footnotesize YONG-JUN ZHANG, BO-QIANG MA and LI-MING YANG}

\address{Department of Physics, Peking University, Beijing 100871,
China}



\maketitle


\begin{abstract}
In this study, the proton is taken as an ensemble of quark-gluon
Fock states. Using the principle of detailed balance, we find
$\bar{d}-\bar{u} \approx 0.124$, which is in surprisingly
agreement with the experimental observation.
\end{abstract}

\section{Hypothesis of Ensemble}

In this talk, we introduce a new work\cite{Zhang01} on the flavor
asymmetry of the nucleon sea as arising from the pure statistical
effect without any parameter. It will be shown that we can
reproduce the flavor asymmetry $\bar{d}-\bar{u} \approx 0.123$,
which is in surprisingly agreement with the experimental
observation, by just using the principle of detailed balance. We
take the proton as an ensemble of a complete set of quark-gluon
Fock states, and these different Fock states can be written as
%
\begin{eqnarray}
|\psi^1\rangle&=&|uud\rangle=|\{uud\},\{0,0,0\}\rangle\\
|\psi^2\rangle&=&|uudg\rangle=|\{uud\},\{0,0,1\}\rangle\\
|\psi^3\rangle&=&|uud\bar{u}u\rangle=|\{uud\},\{1,0,0\}\rangle\\
\cdots\\
|\psi^n\rangle&=&|uud\underbrace{\bar{u}u\cdots\bar{u}u}\limits_i
\underbrace{\bar{d}d\cdots\bar{d}d}\limits_j\underbrace{g\cdots g
}\limits_k\}\rangle=|\{uud\},\{i,j,k\}\rangle,
\end{eqnarray}
where $\{uud\}$ represents the valence quarks of the proton, $i$
is the number of quark-antiquark $u\bar{u}$ pairs, $j$ is the
number of quark-antiquark $d \bar{d}$ pairs, and $k$ is the number
of gluons. Then the density operator of the ensemble is
\begin{eqnarray}
\hat{\rho}&=&\rho_{0,0,0}|uud\rangle\langle
uud|+\rho_{0,0,1}|uudg\rangle\langle uudg|+\cdots \\&=&
\sum_{i,j,k}\rho_{i,j,k}|\{uud\},\{i,j,k\}\rangle\langle
\{uud\},\{ i,j,k\}|,
\end{eqnarray}
where $\rho_{i,j,k}$ is the probability of finding a proton in the
state $|\{uud\},\{i,j,k\}\rangle$ and should satisfy the
normalization condition,
\begin{eqnarray}
\sum_{i,j,k} \rho_{i,j,k}=1. \label{unit}
\end{eqnarray}
It will be shown that $\rho_{i,j,k}$ can be calculated by using
the principle of detailed balance without any parameter.

\section{ The Principle of Detailed Balance}
Given an ensemble of quark-gluon Fock states that has $N_{\rm
A}(N_{\rm A}=N\rho_{\rm A})$ of finding proton in state A and
$N_{\rm B}(N_{\rm A}=N\rho_{\rm A})$ of finding proton in state B,
then the states may change between each others from time to time.
The principle of detailed balance means that the changing between
any two states balance each other. So during a period of time, the
number of events that changed from state A to state B ($n_{{\rm
A}\rightarrow {\rm B}}$) equals the number of events that changed
from state B to state A ($n_{B\rightarrow A}$),
\begin{eqnarray}
n_{{\rm A}\rightarrow {\rm B}}=n_{{\rm B}\rightarrow {\rm
A}}\label{detailed balance}.
\end{eqnarray}
$n_{{\rm A}\rightarrow {\rm B}}$ is proportional to both $N_{\rm
A}$ and the transition probability of A to B ($R_{{\rm
A}\rightarrow {\rm B}}$),
\begin{eqnarray}
n_{{\rm A}\rightarrow {\rm B}}=N_{\rm A}
 R_{{\rm A}\rightarrow {\rm B}}=N \rho_{\rm A} R_{{\rm A}\rightarrow {\rm B}}\label{A-B}.
\end{eqnarray}
We considered about B to A in the same way,
\begin{eqnarray}
n_{{\rm B}\rightarrow {\rm A}}=N_{\rm B}  R_{{\rm B}\rightarrow
{\rm A}}=N \rho_{\rm B} R_{{\rm B}\rightarrow {\rm A}}\label{B-A}.
\end{eqnarray}
Then combining (\ref{detailed balance}), (\ref{A-B}) and
(\ref{B-A}), we have
\begin{eqnarray}
\frac{\rho_{\rm A}}{\rho_{\rm B}}=\frac{R_{{\rm B}\rightarrow {\rm
A}}}{R_{{\rm A}\rightarrow {\rm B}}}\label{ratio}.
\end{eqnarray}
In order to know $\rho_{i,j,k}$, transition probabilities should
be calculated out first.

The transition between two states has two ways: splitting and
recombination.

In the splitting process, the transition probability is
proportional to the number of partons that may split in a certain
state,
\begin{eqnarray}
R_{q\Rightarrow qg}\propto N_q,\ \ \  R_{g\Rightarrow
\bar{q}q}\propto N_g. 
\end{eqnarray}

In the recombination process that involving two kinds of partons,
the transition probability is proportional to both the number of
those two kinds of partons that may recombine in a certain state,
\begin{eqnarray}
R_{qg\Rightarrow q}\propto N_qN_g,\ \ \  R_{\bar{q}q\Rightarrow
g}\propto N_{\bar{q}}N_q.
\end{eqnarray}

For a simple illustration, we neglect the $g\Leftrightarrow gg$
processes because they give small contribution to the flavor
asymmetry\cite{Zhang01}.

We will use symbol
\begin{eqnarray}
|A\rangle{\begin{array}{c}R_{{\rm A}\rightarrow {\rm B}}\\
{\rightleftharpoons} \\ R_{{\rm B}\rightarrow {\rm A}}
\end{array}}|B\rangle
\end{eqnarray} to present our calculations about
transition probabilities.

(1) In the translation process involving the creation or
annihilation of gluon, only $q\Leftrightarrow qg$ will be
considered while $g\Leftrightarrow gg$ neglected.
So we have
\begin{eqnarray} |\{uud\},\{i,j,k-1\}\rangle{\begin{array}{c}3+2i+2j\\
{\rightleftharpoons}
\\ (3+2i+2j)k \end{array}}|\{uud\},\{i,j,k\}\rangle.
\end{eqnarray}
Using formula (\ref{ratio}), we obtain a recursion formula,
\begin{eqnarray}
\frac{\rho_{i,j,k}}{\rho_{i,j,k-1}}=\frac{1}{k}.
\end{eqnarray}
We can further get a more general formula,
\begin{eqnarray}
\frac{\rho_{i,j,k}}{\rho_{i,j,0}}=\frac{1}{k!}\label{g-gg},
\end{eqnarray}
 and we have
$\rho_{i,j,1}=\rho_{i,j,0}$, which we will used later.

(2) In the translation process involving the creation or
annihilation of a pair of $\bar{u}u$, we have
\begin{eqnarray} |\{uud\},\{i-1,j,1\}\rangle{\begin{array}{c}1\\
{\rightleftharpoons}
\\ i(i+2) \end{array}}|\{uud\},\{i,j,0\}\rangle.
\end{eqnarray} Using formula (\ref{ratio}), we obtain a recursion formula,
\begin{eqnarray}
\frac{\rho_{i,j,0}}{\rho_{i-1,j,1}}=\frac{1}{i(i+2)}.
\end{eqnarray} Using relation
$\rho_{i,j,1}=\rho_{i,j,0}$, we get
\begin{eqnarray}
\frac{\rho_{i,j,0}}{\rho_{i-1,j,0}}=\frac{1}{i(i+2)},\\
\frac{\rho_{i,j,0}}{\rho_{0,j,0}}=\frac{2}{i!(i+2)!}\label{g-uu}.
\end{eqnarray}

(3) In the translation process involving the creation or
annihilation of a pair of $\bar{d}d$, we have
\begin{eqnarray} |\{uud\},\{i,j-1,1\}\rangle{\begin{array}{c}1\\
{\rightleftharpoons}
\\ j(j+1) \end{array}}|\{uud\},\{i,j,0\}\rangle.
\end{eqnarray} Using formula (\ref{ratio}), we obtain a recursion formula,
\begin{eqnarray}
\frac{\rho_{i,j,0}}{\rho_{i,j-1,1}}=\frac{1}{j(j+1)}.
\end{eqnarray} Using relation
$\rho_{i,j,1}=\rho_{i,j,0}$, we get
\begin{eqnarray}
\frac{\rho_{i,j,0}}{\rho_{i,j-1,0}}=\frac{1}{j(j+1)},\\
\frac{\rho_{i,j,0}}{\rho_{i,0,0}}=\frac{1}{j!(j+1)!}\label{g-dd}.
\end{eqnarray}
Combining (\ref{g-gg}), (\ref{g-uu}), and (\ref{g-dd}), we obtain
the general formula
\begin{equation}
\frac{\rho_{i,j,k}}{\rho_{0,0,0}}=\frac{2}{i!(i+2)!j!(j+1)!k!}\label{aijk}.
\end{equation}
 From this result and the normalization condition (\ref{unit}), all
$\rho_{i,j,k}$ can be calculated out as shown in Table 1. From
Table 1, we get the numbers of intrinsic gluons and sea quarks of
the proton,
\begin{eqnarray}
\bar{u}=\sum_{i,j,k}
i\rho_{i,j,k}=0.308,\\ \bar{d}=\sum_{i,j,k} j\rho_{i,j,k}=0.432,\\
 g=\sum_{i,j,k} k\rho_{i,j,k}=0.997,\\
 \bar{d}-\bar{u}=0.124.
\end{eqnarray}

The flavor sea-quark asymmetry $\bar{d}-\bar{u}$ can be checked by
experiments directly because its $Q^2$ dependence is small. It is
a surprise that our result is in excellent agreement with the
recent experimental result $\bar{d}-\bar{u}=0.118 \pm 0.012$
\cite{E866}. This good agreement indicates that the principle of
detailed balance plays an essential role in the structure of
proton. We also give a complete set of Fock states for the proton,
with the probability of finding each Fock state calculated without
any parameter, as shown in Table~1.

\begin{table}[htbp]
\ttbl{30pc}{The probabilities, $\rho_{i,j,k}$,
  of finding the quark-gluon Fock states of the
  proton.}
{\begin{tabular}{lllcccccc}\\
\multicolumn{9}{c}{} \\[6pt]\hline
      {i}&{j}&{$|\{uud\},\{i,j,0\}\rangle$}&{$\rho_{i,j,0}$}&{$\rho_{i,j,1}$}&{$\rho_{i,j,2}$}&{$\rho_{i,j,3}$}
        &{$\rho_{i,j,4}$}&{$\cdots$}\\ \hline
        0&0&$|uud\rangle$         & 0.167849&0.167849&0.083924&0.027975&0.006994&$\cdots$\\
        1&0&$|uud\bar{u}u\rangle$ &0.055950&0.055950&0.027975&0.009325&0.002331&$\cdots$\\
        0&1&$|uud\bar{d}d\rangle$ &0.083924&0.083924&0.041962&0.013987&0.003497&$\cdots$\\
        1&1&$|uud\bar{u}u\bar{d}d\rangle$ &0.027975&0.027975&0.013987&0.004662&0.001166&$\cdots$\\
        0&2&$|uud\bar{d}d\bar{d}d\rangle$ &0.013987&0.013987&0.006994&0.002331&0.000583&$\cdots$\\
        2&0&$|uud\bar{u}u\bar{u}u\rangle$ &0.006994&0.006994&0.003497&0.001166&0.000291&$\cdots$\\
        1&2&$|uud\bar{u}u\bar{d}d\bar{d}d\rangle$ &0.004662&0.004662&0.002331&0.000777&0.000194&$\cdots$\\
        2&1&$|uud\bar{u}u\bar{u}u\bar{d}d\rangle$ &0.003497&0.003497&0.001748&0.000583&0.000146&$\cdots$\\
        0&3&$|uud\bar{d}d\bar{d}d\bar{d}d\rangle$ &0.001166&0.001166&0.000583&0.000194&0.000049&$\cdots$\\
        3&0&$|uud\bar{u}u\bar{u}u\bar{u}u\rangle$ &0.000466&0.000466&0.000233&0.000078&0.000019&$\cdots$\\
         $\cdots$&$\cdots$&$\cdots$&$\cdots$&$\cdots$&$\cdots$&$\cdots$&$\cdots$&$\cdots$\\
\hline
\end{tabular}}
\end{table}



\begin{thebibliography}{0}
\bibitem{Zhang01}
Y.-J. Zhang, B. Zhang and B.-Q. Ma, {\it Phys.\ Lett.\ } {\bf
B523}, 260 (2001). There is one misprint in formula (11), in which
``3" should be ``5".
\bibitem{E866}
FNAL E866/NuSea Collaboration, R.S. Towell {\it et al.}, {\it
Phys. Rev. } {\bf D64}, 052002 (2001). For a recent review, see,
e.g.,
G.~T.~Garvey and J.~C.~Peng,
{\it Prog.\ Part.\ Nucl.\ Phys.\ }  {\bf 47}, 203 (2001).



\end{thebibliography}
\end{document}